\documentstyle[preprint,aps,epsf]{revtex}
\begin{document}
\draft         
\preprint{\font\fortssbx=cmssbx10 scaled \magstep2
\hfill$\vcenter{\hbox{\bf CERN-TH/96-123}
		\hbox{\bf IFT--P.013/96}
}$}
\title{Excited Fermion Contribution to Z$^0$ Physics at One Loop}
\author{M.\ C.\ Gonzalez-Garcia}
\address{Theory Division, CERN,
CH-1211 Geneva 23, Switzerland.}
\author{ S.\ F.\ Novaes }
\address{Instituto de F\'{\i}sica Corpuscular -- IFIC/C.S.I.C. \\
Dept.\ de F\'{\i}sica Te\'orica, Universidad de Valencia \\
46100 Burjassot, Valencia, Spain\\
and \\
Instituto de F\'{\i}sica Te\'orica, 
Universidade  Estadual Paulista, \\  
Rua Pamplona 145, CEP 01405-900 S\~ao Paulo, Brazil.}
%
\maketitle
\widetext
\begin{abstract}
We investigate the effects induced by excited leptons at the one-loop 
level in the observables measured on the $Z$ peak at LEP.
Using a general effective Lagrangian approach to describe the
couplings of the excited leptons, we compute their contributions
to both oblique parameters and $Z$ partial widths. Our results
show that the new effects are comparable to the present
experimental sensitivity, but they do not lead to a significant
improvement on the available constraints on the couplings and
masses of these states.
\end{abstract}
\noindent
{\bf CERN-TH/96-123} \\
\noindent
{\bf May 1996}\\
\newpage


\section{Introduction}
\label{sec:1}

The standard model of electroweak interactions (SM) is not able
to give a satisfactory explanation to family repetition and to
the complex pattern of  the fermion masses. One expects  a
substantial improvement  in the  understanding of these problems
when considering  an underlying fermionic substructure where the
usual fermions share some constituents (preons) \cite{comp}. In
this sense, the SM would be just the low-energy limit of a more
fundamental theory, being valid only at energies below the
compositeness mass  scale $\Lambda$.  

One of the most unambiguous predictions of the composite models
is the  existence of an excited lepton state for each known
lepton. Unfortunately, we do not yet have a satisfactory model
that could reproduce  the whole family spectrum. In view of the
lack of a unique predictive theory, a model-independent
phenomenological analysis of the effects of fermion compositeness
seems the most appealing approach.  On this ground, we can
employ the effective Lagrangian techniques to  describe the
physics of these excited states below the compositeness scale. 

This approach has been employed in several phenomenological
studies that  analysed the expected signatures of these excited
fermions in $pp$ \cite{kuhn,pp}, $e^+e^-$
\cite{kuhn,hag,ele:pos,lep:hera,e:gam}, and $ep$
\cite{hag,lep:hera} collisions at high energies. On the
experimental side, several searches for these particles have been
carried out, including those at the CERN Large Electron--Positron
Collider (LEP) \cite{rev:lep} and at HERA \cite{hera}. At LEP,
the experiments at the $Z$ pole excluded the existence of excited
spin--$\frac{1}{2}$ fermions with mass up to $46$ GeV from the
pair production search ($e^+ e^- \to \ell^\ast \ell^\ast$),  and
up to $90$ GeV from direct single production ($e^+ e^- \to \ell
\ell^\ast$) for a scale of compositeness $\Lambda < 2.5$ TeV
\cite{rev:lep}. Very recent results from the L3 Collaboration
\cite{lep:130}, at centre-of-mass energies of 130--140 GeV,
determined the lower mass limits at 95\% C.L.\ of 64.7 GeV for
the excited electrons, and roughly $\Lambda \geq 1.4$ TeV  for
$90 \leq M_{e^\ast} \leq 130$ GeV. The experiments at the DESY
$ep$ collider HERA also searched for resonances in the $e\gamma$,
$\nu W$, and $eZ$ systems \cite{hera,zeus}; however the LEP
bounds on excited leptons couplings are about one order of
magnitude more stringent in the mass region below the $Z$ mass. 

In spite of the failure of all the direct searches for
compositeness, we could expect that the next generation of
accelerators, working  at higher centre-of-mass energies, would
be able to obtain a direct evidence of the  existence of these
composite states. On the other hand, an important source of
indirect information about new particles and interactions is the
precise measurement of the electroweak parameters done at LEP.
Virtual effects of these new states can alter the SM
predictions for some of these parameters and the comparison with
the experimental data can impose bounds on their masses and
couplings.

In this work we investigate the one-loop effects of excited
leptons in the observables measured on the $Z$ peak at LEP. Using
an effective Lagrangian in terms of dimension five operators to
describe the couplings of the  excited leptons, we compute their
contribution to both oblique and vertex corrections to the
electroweak parameters. Our results show that the new effects are
comparable to the present experimental sensitivity, but they are
only able to constrain very marginally the model parameters
beyond the present limits from direct searches.

The outline of the paper is as follows. In section \ref{sec:eff},
we introduce the effective Lagrangian describing the couplings of
the excited leptons. Section \ref{sec:ana} contains the relevant
analytical expressions for the one-loop corrections induced by
the excited leptons. Our results and their respective discussion are
given in Section \ref{sec:disc}. This paper is supplemented with 
Appendix \ref{pave} where we list all the relevant
Passarino--Veltman functions.

\section{Effective Interactions}
\label{sec:eff}

In order to reduce the number of free parameters in a general
effective Lagrangian, we concentrate here in a specific model,
following the formulation of Hagiwara {\it et al.\/} \cite{hag}.
This particular model has been used by several experimental
collaborations as a guideline to the search of composite states.
We consider excited fermionic states with spin and isospin
$\frac{1}{2}$, and we assume that the excited fermions acquire
their masses before the $SU(2) \times U(1)$ breaking, so that
both left-handed and right-handed states belong to weak
isodoublets. We introduce the weak doublets, with hypercharge $Y
= -1$, for the usual left-handed fermion ($\psi_L$) and for the
excited fermions ($\Psi^\ast$),
\[
\psi_L = \left(\begin{array}{c}
\nu \\
e
\end{array}
\right)_L \;\; , \; \text{and} \;\;\;\;\;\;
\Psi^\ast = \left(\begin{array}{c}
N \\
E
\end{array}
\right) \; ,
\]
The most general dimension-five effective Lagrangian describing
the coupling of the excited fermions to the usual fermions, which
is  $SU(2) \times U(1)$ invariant and CP conserving can be
written as \cite{hag}
\begin{equation}
{\cal L}_{Ff} = - \frac{1}{2 \Lambda} {\bar \Psi^\ast} \sigma^{\mu\nu}
\left(g f_2 \frac{\tau^i}{2} W^i_{\mu\nu} + 
      g' f_1 \frac{Y}{2} B_{\mu\nu}\right) \psi_L 
+ \; \text{h. c.} \; ,
\label{l:eu:0}
\end{equation}
where $f_2$ and $f_1$ are weight factors associated to the
$SU(2)$ and $U(1)$ coupling constants, with $\Lambda$ being the
compositeness scale, and  $\sigma_{\mu\nu} = (i/2)[\gamma_\mu,
\gamma_\nu]$. $g$ and $g'$ are the gauge coupling constants of
$SU(2)$ and $U(1)$ respectively. At tree-level they can  be
expressed in terms  of the electric charge, $e$, and the Weinberg
angle, $\theta_W$, as  $g=e/\sin\theta_W$ and
$g'=e/\cos\theta_W$. We will assume a pure left-handed structure
for these couplings in order to comply with  the strong bounds
coming from the measurement of the anomalous magnetic  moment of
leptons \cite{g-2}. 

In terms of the physical fields, the Lagrangian (\ref{l:eu:0})
becomes
\begin{equation}
{\cal L}_{Ff} = - \sum_{V=\gamma,Z,W} 
C_{VFf} \bar{F} \sigma^{\mu\nu} (1 - \gamma_5) f \partial_\mu V_\nu 
- i \sum_{V=\gamma,Z} D_{VFf} \bar{F} \sigma^{\mu\nu} (1 - \gamma_5) f
W_\mu V_\nu + \; \text{h. c.} \; ,
\label{l:eu}
\end{equation}
where $F = N, E$, and $f = \nu, e$. The non-abelian structure of
(\ref{l:eu:0}) gives rise to a contact quartic interaction,
such as the second term in the r.h.s. of Eq.\ (\ref{l:eu}). In this
equation, we have omitted terms containing two $W$ bosons, which
do not play any role in our calculations. $C_{VFf}$ is the
coupling of the vector boson with the different kinds of fermions,
\begin{equation}
\begin{array}{ll}
C_{\gamma E e} =  - \frac{\displaystyle e}{\displaystyle 4 \Lambda} 
(f_2 + f_1) &
\;\; , \;\;\;\;
C_{\gamma N \nu} =  
\frac{\displaystyle e}{\displaystyle 4 \Lambda} (f_2 - f_1)\\
C_{Z E e} =  
- \frac{\displaystyle e}{\displaystyle 4 \Lambda} 
(f_2 \cot\theta_W - f_1 \tan\theta_W) &
\;\; , \;\;\;\;
C_{Z N \nu} =  
\frac{\displaystyle e}{\displaystyle 4 \Lambda} 
(f_2 \cot\theta_W + f_1 \tan\theta_W)\\
C_{W E \nu} =   C_{W N e} = 
\frac{\displaystyle e}{\displaystyle 2 \sqrt{2} \sin\theta_W \Lambda} f_2\; , & 
\end{array}
\label{CV}
\end{equation}
and the quartic interaction coupling constant, $D_{VFf}$, is given by
\begin{equation}
\begin{array}{l}
D_{\gamma E \nu} = - D_{\gamma N e} =  
 \frac{\displaystyle e^2 \sqrt{2}}{\displaystyle 4 \sin\theta_W \Lambda} f_2\\
D_{Z E \nu} = - D_{Z N e} = 
 \frac{\displaystyle e^2 \sqrt{2} \cos\theta_W}
{\displaystyle 4 \sin^2\theta_W \Lambda} f_2\; .
\end{array}
\label{DV}
\end{equation}

The coupling of gauge bosons to excited leptons can be described
by the $SU(2) \times U(1)$ invariant and CP conserving, effective
Langragian,
\begin{equation}
{\cal L}_{FF} = 
- \bar{\Psi^\ast} \Biggl[ \left(g \frac{\tau^i}{2} \gamma^\mu W^i_{\mu} + 
g' \frac{Y}{2}  \gamma^\mu B_\mu \right) + 
\left(\frac{g \kappa_2}{2 \Lambda} \frac{\tau^i}{2} \sigma^{\mu\nu}
\partial_\mu W^i_{\nu} + 
\frac{g' \kappa_1}{2 \Lambda}  \frac{Y}{2} 
\sigma^{\mu\nu}\partial_\mu  B_\nu \right) 
\Biggr] \Psi^\ast
\label{l:ee:0}
\end{equation}
In terms of the physical fields, this can be written as, 
\begin{equation}
{\cal L}_{FF} = - \sum_{V=\gamma,Z,W} \bar{F} 
(A_{VFF} \gamma^\mu V_\mu + K_{VFF} \sigma^{\mu\nu} \partial_\mu V_\nu) F\; .
\label{l:ee}
\end{equation}

Since we have assumed that the left- and right-handed excited
leptons have the same quantum numbers under the standard gauge
group, the dimension-four piece in (\ref{l:ee}) is taken
vector-like. $A_{VFF}$ is given by
\begin{equation}
\begin{array}{ll}
A_{\gamma EE} =  - e &
\;\; , \;\;\;\;
A_{\gamma NN} =   0 \\
A_{ZEE} = e \frac{\displaystyle (2 \sin^2\theta_W - 1)}
{\displaystyle 2  \sin\theta_W  \cos\theta_W} &
\;\; , \;\;\;\; 
A_{ZNN} =  \frac{\displaystyle e}{\displaystyle 2  \sin\theta_W  \cos\theta_W}
\\
A_{WEN} = \frac{\displaystyle e}{\displaystyle \sqrt{2} \sin\theta_W} &
\end{array}
\label{AV}
\end{equation}
and $K_{VFF}$ is given by
\begin{equation}
\begin{array}{ll}
K_{\gamma EE} =
- \frac{\displaystyle e}{\displaystyle 4 \Lambda} (\kappa_2 + \kappa_1) & 
\;\; , \;\;\;\;
K_{\gamma NN} =  
\frac{\displaystyle e}{\displaystyle 4 \Lambda} (\kappa_2 - \kappa_1)\\
K_{Z EE} = 
- \frac{\displaystyle e}{\displaystyle 4 \Lambda}
(\kappa_2\cot\theta_W -\kappa_1\tan\theta_W)  &
\;\; , \;\;\;\;
K_{Z NN} =  
\frac{\displaystyle e}{\displaystyle 4 \Lambda} 
(\kappa_2 \cot\theta_W + \kappa_1\tan\theta_W)
\\ 
K_{WEN} =  
\frac{\displaystyle e}{\displaystyle 2 \Lambda} 
\frac{\displaystyle \kappa_2}{\displaystyle \sqrt{2} \sin\theta_W}.
\end{array}
\label{KV}
\end{equation}

It is important to notice that the phenomenological model for the
excited fermions described by the Lagrangians (\ref{l:eu}) and
(\ref{l:ee}) has been extensively used by several experimental
collaborations \cite{rev:lep,hera,zeus,lep:130} to search for
excited states. Therefore the results presented in this paper can
be directly compared with the bounds on the excited fermion mass
and compositeness scale obtained by these collaborations.

\section{Analytical Expressions}
\label{sec:ana}

In this work we employed the on-shell-renormalization scheme,
adopting the conventions of Ref.\ \cite{hollik}. We used as
inputs the fermion masses, $G_F$, $\alpha$, and the $Z$-boson
mass. The electroweak mixing angle is a derived quantity  defined
through $\sin^2 \theta_W = s_W^2 \equiv 1 - M^2_W / M^2_Z$.

As a general procedure to evaluate the virtual contributions of
the excited states, with couplings described by (\ref{l:eu}) and
(\ref{l:ee}), we evaluated the loops in $D = 4 - 2 \epsilon$
dimensions using the dimension regularization method
\cite{reg:dim} which is a gauge-invariant regularization
procedure, and we adopted the unitary gauge to perform the
calculations. We identified the poles at $D=4$ ($\epsilon=0$) and
$D=2$ ($\epsilon= 1$) with the logarithmic and quadratic
dependence on the scale $\Lambda$ \cite{zep}. The finite part of
the loop is given by 
\[
L_{\text{finite}} = \lim_{\epsilon \to 0} \left[ L(\epsilon) - 
R_0 \left( \frac{1}{\epsilon} - \gamma_E + \log 4\pi + 1 \right)
- R_1 \left( \frac{1}{\epsilon - 1} + 1  \right) 
\right]
\]
where $R_{0(1)}$ are the residues of the poles at $\epsilon =
0 (1)$. The final result is written as
\[
L = L_{\text{finite}} + R_0 \log \left(\frac{\Lambda^2}{\mu^2}\right) + 
R_1 \frac{\Lambda^2}{4 \pi \mu^2}
\]

In order to compute the loops in $D$ dimensions in terms of 
the Passarino--Veltman scalar one-loop functions (see Appendix
\ref{pave}), we used the Mathematica package FeynCalc
\cite{feyn}. The output of FeynCalc, in the case of the
two-point functions, was checked against the results obtained by
a direct analytical calculation. 

Close to the $Z$ resonance, the physics can be summarized by the
effective neutral current
\begin{equation}
J_\mu =  \left ( \sqrt{2} G_\mu M_Z^2 \rho_f 
\right )^{1/2} \left [ \left ( I_3^f - 2 Q^f s_W^2 \kappa_f \right )
\gamma_\mu - I_3^f \gamma_\mu \gamma_5 \right ] \; ,
\label{form:nc}
\end{equation}
where $Q^f$ ($I_3^f$) is the fermion electric charge (third
component of weak isospin), and $G_\mu$ is the Fermi coupling
constant measured via the muon lifetime. The form factors
$\rho_f$ and $\kappa_f$  have universal contributions, {\em i.e.}
independent of the fermion species, as well as non-universal
parts,
\begin{eqnarray}
 \rho_f   & = &  1 + \Delta \rho_{\text{univ}} + 
\Delta \rho_{\text{non}} \; , \\
\kappa_f  & = &  1 + \Delta \kappa_{\text{univ}} + 
\Delta \kappa_{\text{non}} \; .
\end{eqnarray}

Excited leptons can affect the physics at the $Z$ pole through
their contributions to both universal and non-universal
corrections. The universal contributions can be expressed in
terms of the unrenormalized vector boson self-energies. Defining
the transverse part of vacuum polarization amplitudes between
the vector boson $V_1 - V_2$, $\Pi^{V_1V_2}_{\mu\nu}(q^2)$, as
\[
\Pi^{V_1V_2}_{\mu\nu}(q^2) \equiv \; g_{\mu\nu} \; \Sigma^{V_1V_2}(q^2)
\]
where $V_{1,2} = \gamma$, $W$, and $Z$, we can write

\begin{equation}
\begin{array}{l}
\Delta \rho^{\text{ex}}_{\text{univ}}(s)  =   
-\frac{\displaystyle \Sigma^{ZZ}_{\text{ex}}(s)-\Sigma^{ZZ}_{\text{ex}}(z)}
{\displaystyle  s-z} 
+\frac{\displaystyle \Sigma^{ZZ}_{\text{ex}}(z)}{\displaystyle z}
-\frac{\displaystyle \Sigma^{WW}_{\text{ex}}(0)}{\displaystyle w } + 
2 \frac{\displaystyle s_W}{\displaystyle c_W}
\frac{\displaystyle \Sigma^{\gamma Z}_{\text{ex}}(0)}
{\displaystyle  z}
\; ,\\
\Delta \kappa^{\text{ex}}_{\text{univ}}   =   
\frac{\displaystyle c_W}{\displaystyle s_W} 
\frac{\displaystyle \Sigma^{\gamma Z}_{\text{ex}}(z)}{\displaystyle z}
+ \frac{\displaystyle c_W}{\displaystyle s_W}~ 
\frac{\displaystyle \Sigma^{\gamma Z}_{\text{ex}}(0)}{\displaystyle z}
+\frac{\displaystyle c_W^2}{\displaystyle s_W^2} 
\left[ \frac{\displaystyle \Sigma_{\text{ex}}^{ZZ} (z)}{\displaystyle z}-
\frac{\displaystyle \Sigma_{\text{ex}}^{WW}(w)}{\displaystyle w}\right]\; ,\\
\Delta r^{\text{ex}}_{\text{univ}} =  
\Sigma_{\text{ex}}^{\prime \gamma\gamma}(0)-
\frac{\displaystyle c_W^2}{\displaystyle s_W^2}\left(
\frac{\displaystyle \Sigma_{\text{ex}}^{ZZ}(z)}{\displaystyle z} 
-\frac{\displaystyle \Sigma_{\text{ex}}^{WW}(w)}{\displaystyle w}\right)
+ \frac{\displaystyle\Sigma_{\text{ex}}^{WW}(0)- 
\Sigma_{\text{ex}}^{WW}(w)}{\displaystyle w}
-2\frac{\displaystyle c_W}{\displaystyle s_W}
\frac{\displaystyle \Sigma_{\text{ex}}^{\gamma Z}(0)}{\displaystyle z}
\; ,
\end{array}
\end{equation}
where $w(z) = M^2_{W(Z)}$, $s_W(c_W)=\sin(\cos) \theta_W$ and
$\Sigma^{\; \prime} = d\Sigma/dq^2$.

\begin{figure}
\epsfxsize=8cm
\begin{center}
\leavevmode \epsfbox{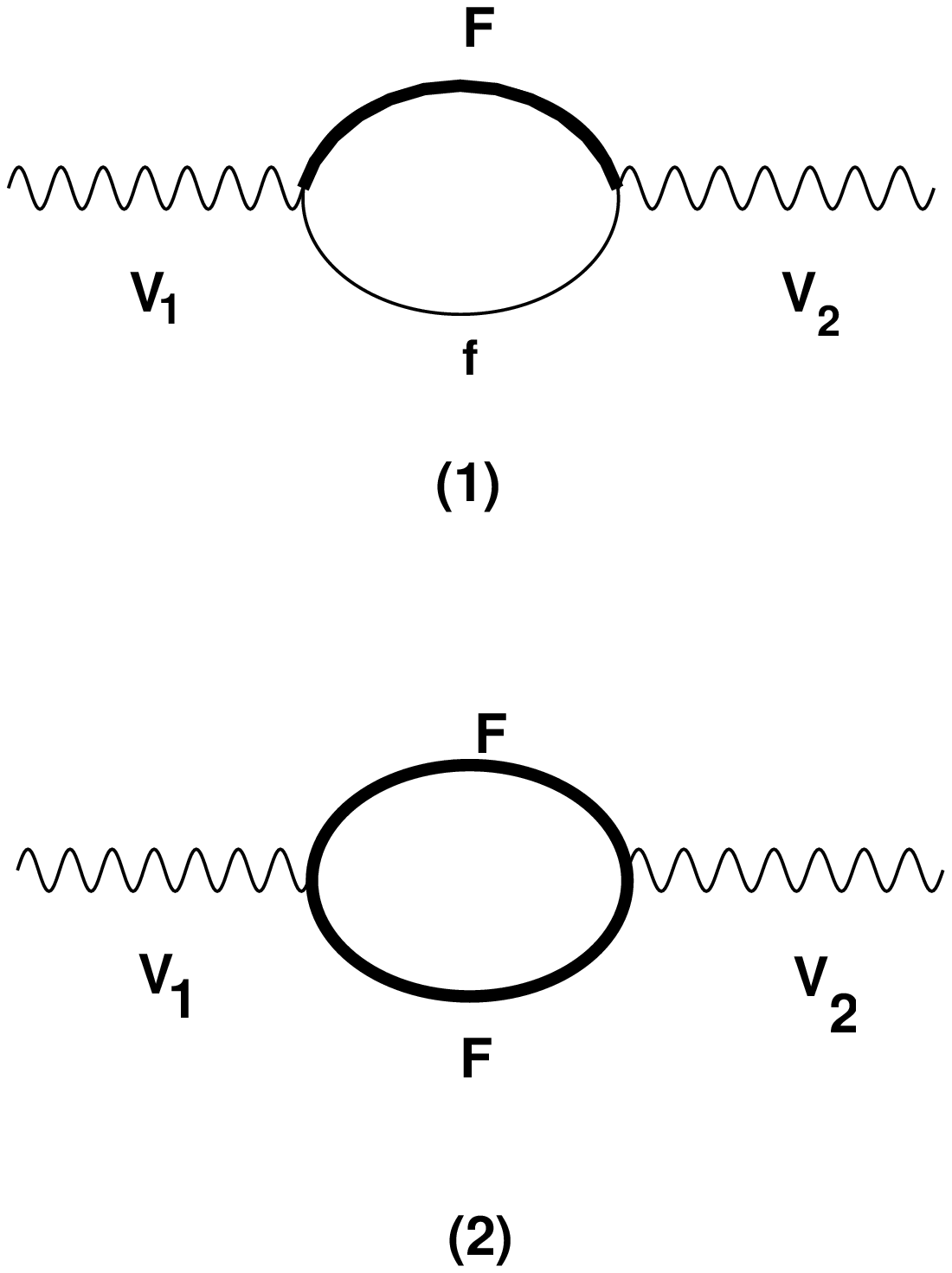}
\end{center}
\caption{Feynman diagrams leading to  contribution of the excited leptons 
to the two-point functions}
\label{fig:1}
\end{figure}

The diagrams with excited lepton  contributions to the
self-energies are shown in Fig.\ \ref{fig:1}.  The final result
for the transverse part of vacuum polarization
$\Sigma^{V_1V_2}_{Ff}$ contribution coming from the loop of an
excited fermion with mass $M$ and an ordinary massless fermion is 
\begin{equation}
\begin{array}{ll}
\Sigma^{V_1 V_2}_{Ff} =&  
 \frac{\displaystyle 1}{\displaystyle 12 \pi^2} C_{V_1Ff} C_{V_2Ff}
\;
\Biggl\{ 6 q^2 \Lambda^2 + 
q^4 \log\frac{\displaystyle \Lambda^2}{\displaystyle M^2} \\
&- 2 q^2 M^2 -\frac{\displaystyle q^4}{\displaystyle 3} 
+ M^2(2M^2-q^2) + (M^2 - q^2)(2M^2 + q^2)\\
&\times 
\left[-2 +
\left(1-\frac{\displaystyle M^2}{\displaystyle q^2}\right)
\log\left(1-\frac{\displaystyle q^2}{\displaystyle M^2} \right)
\right]
\Biggr\}
\end{array}
\label{pi:eu:final}
\end{equation}
where $V_{1(2)}$ refers to the initial (final) vector boson,
and the constants $C_{VFf}$ are defined in (\ref{CV}) for the
different vector bosons and fermions.

For the vacuum polarization, $\Sigma^{V_1V_2}_{FF}$,  coming from
the loop of two excited fermions with mass $M$, we obtain:
\begin{equation}
\begin{array}{ll}
\Sigma^{V_1 V_2}_{FF} =&  
\frac{\displaystyle 1}{\displaystyle 24\pi^2}  \; 
\frac{\displaystyle q^2}{\displaystyle M^2}
\Biggl\{6 K_{V_1FF} K_{V_2FF} M^2 \Lambda^2 
+ \Bigl[ 2 A_{V_1FF} A_{V_2FF} 
\\ 
&  + 6 (A_{V_1FF} K_{V_2FF} + A_{V_2FF} K_{V_1FF}) M 
+ 3 K_{V_1FF} K_{V_2FF} 
\left(\frac{\displaystyle q^2}{\displaystyle 3} + 2 M^2\right)
\Bigr ] M^2 
\log \frac{\displaystyle \Lambda^2}{\displaystyle M^2}\\
&+ 4 A_{V_1FF} A_{V_2FF}M^2
\left( \frac{\displaystyle 1}{\displaystyle 3}+ 
\frac{\displaystyle 2 M^2}{\displaystyle q^2}\right)
+ 6 (A_{V_1FF} K_{V_2FF} + A_{V_2FF}
K_{V_1FF}) M^3  \\
&  + K_{V_1FF} K_{V_2FF} M^2
\left(\frac{\displaystyle 5 q^2}{\displaystyle 3} + 4 M^2\right) \\
&- 2 \frac{\displaystyle (4 M^2 -q^2)^{1/2}}{\displaystyle q}
\arctan\left[\frac{\displaystyle q}{\displaystyle (4 M^2 - q^2)^{1/2}}\right]
\left[
2 A_{V_1FF} A_{V_2FF}M^2
\left( 1+ 
\frac{\displaystyle 2 M^2}{\displaystyle q^2}\right)
\right. \\
& \left. + 6 (A_{V_1FF} K_{V_2FF} + A_{V_2FF}
K_{V_1FF}) M^3 
+ K_{V_1FF} K_{V_2FF} M^2
\left(q^2+ 8 M^2\right) 
\right]
\Biggr\}
\end{array}
\label{pi:ee:final}
\end{equation}

For the purpose of illustration, we derived  approximate
expressions for the excited fermion contribution to the two-point
functions, $\Sigma^{V_1V_2}_{\text{ex}}$, in the large-$M$ limit.
For $R_Q \equiv q^2/M^2 \ll 1$, we obtain
\begin{equation}
\begin{array}{ll}
\Sigma^{V_1V_2}_{\text{ex}}=& \Sigma^{V_1V_2}_{\text{Ff}}+
\Sigma^{V_1V_2}_{\text{FF}} \\
 = &
\frac{\displaystyle M^2}{\displaystyle 12\pi^2} R_Q 
\Biggl\{3\Lambda^2 ( 2C_{V_1Ff} C_{V_2Ff}+K_{V_1FF} K_{V_2FF})
-A_{V_1FF} A_{V_2FF}\\
&  -3 (A_{V_1FF} K_{V_2FF} + A_{V_2FF} K_{V_1FF}) M 
- 6 K_{V_1FF} K_{V_2FF}M^2 - 3 C_{V_1Ff} C_{V_2Ff}M^2
 \\
& + \left[ A_{V_1FF} A_{V_2FF} + 3 (A_{V_1FF} K_{V_2FF} + 
A_{V_2FF} K_{V_1FF}) M 
+ 3 K_{V_1FF} K_{V_2FF} M^2\right] 
\log \frac{\displaystyle \Lambda^2}{\displaystyle M^2}\Biggr\}
\end{array}
\end{equation} 
We obtain in this approximation the following expressions for the 
universal corrections,
\begin{equation}
\begin{array}{rl}
\Delta\rho(z)=&\frac{\displaystyle \alpha}{\displaystyle 720 c_W^2 s_W^2 \pi}
 \, R_Z\, \left(c_W^4 + s_W^4 \right)
       \Biggl[ -24 - 60\,k\, \sqrt{R_L} - 50\, f^2 \,  R_L - 
        15\, k^2 \, R_L 
\\
& + 60\, f^2 \, R_L \,\log R_L + 
        15\, k^2 \, R_L \,\log R_L \Biggr] \\
\Delta\kappa= & -\frac{\displaystyle c_W^4}{\displaystyle c_W^4+s_W^4}
\Delta\rho(z)\\
\Delta r= & \frac{\displaystyle c_W^2}{\displaystyle c_W^4+s_W^4}
\Delta\rho(z)\\
\end{array}
\label{uniapprox}
\end{equation}
where $R_Z \equiv M_Z^2/M^2$ and $R_L \equiv M^2/\Lambda^2$. For
the sake of simplicity, we have assumed that $f_1 = f_2=f$ and
$k_1=k_2=k$. 

Since we are considering non-renormalizable dimension-five
operators the loops should, in principle, present poles at $D =
2$ that would generate terms that are finite when $\Lambda \to
\infty$. However, we are restricting ourselves to $SU(2)\times
U(1)$ gauge invariant operators, and the final results for the
physical observables behave, at most, like
$\log\Lambda^2/\Lambda^2$, after using the SM counterterms [see
Eq.\ (\ref{uniapprox})]. Also,  it is straightforward to verify
that the new physics decouples as the new contributions in Eq.\
(\ref{uniapprox}) vanish in the limit $R_Z\rightarrow 0$ for
fixed $R_L$.

\begin{figure}
\protect
\epsfxsize=15cm
\begin{center}
\leavevmode \epsfbox{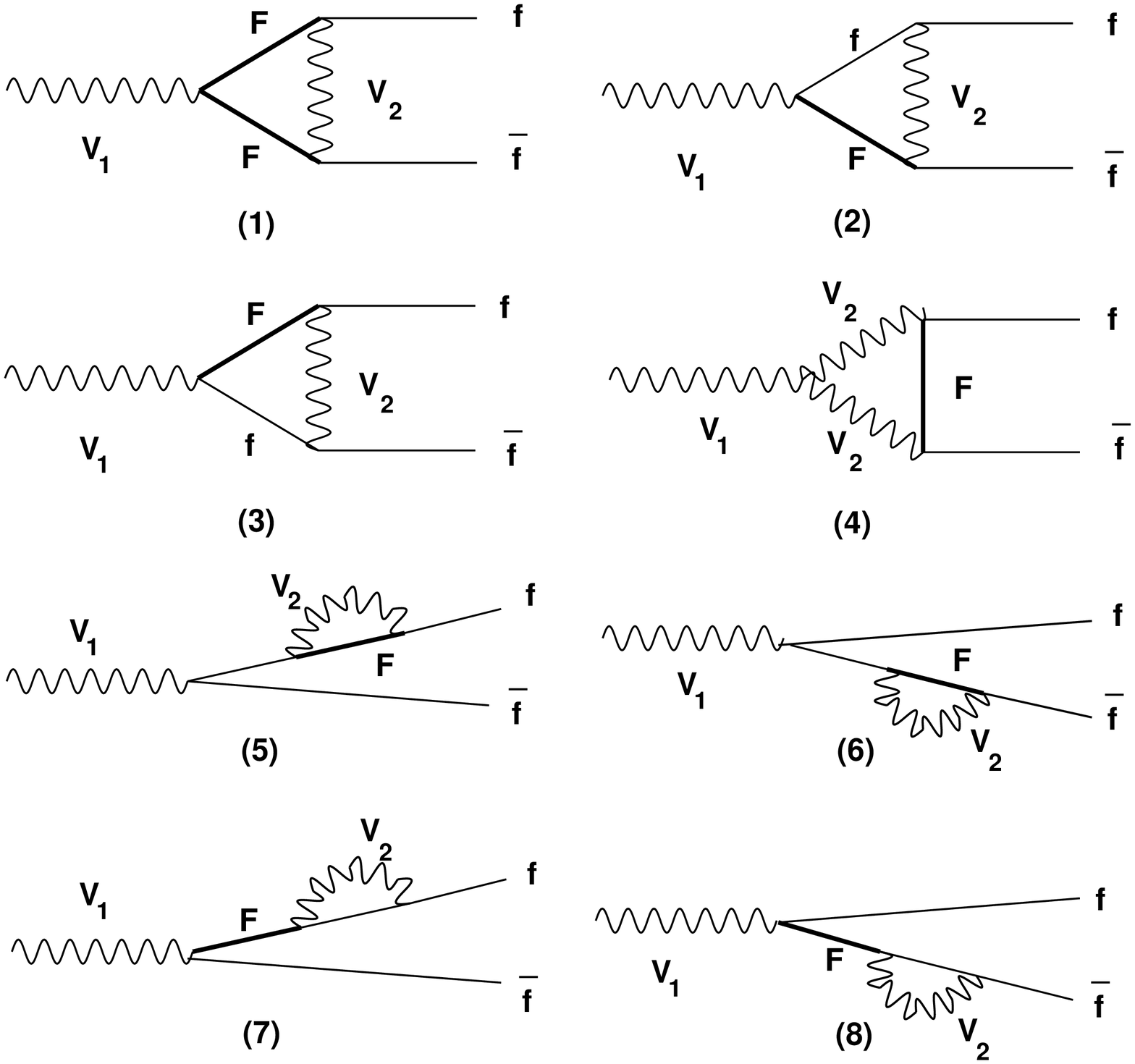}
\end{center}
\epsfxsize=15cm
\begin{center}
\leavevmode \epsfbox{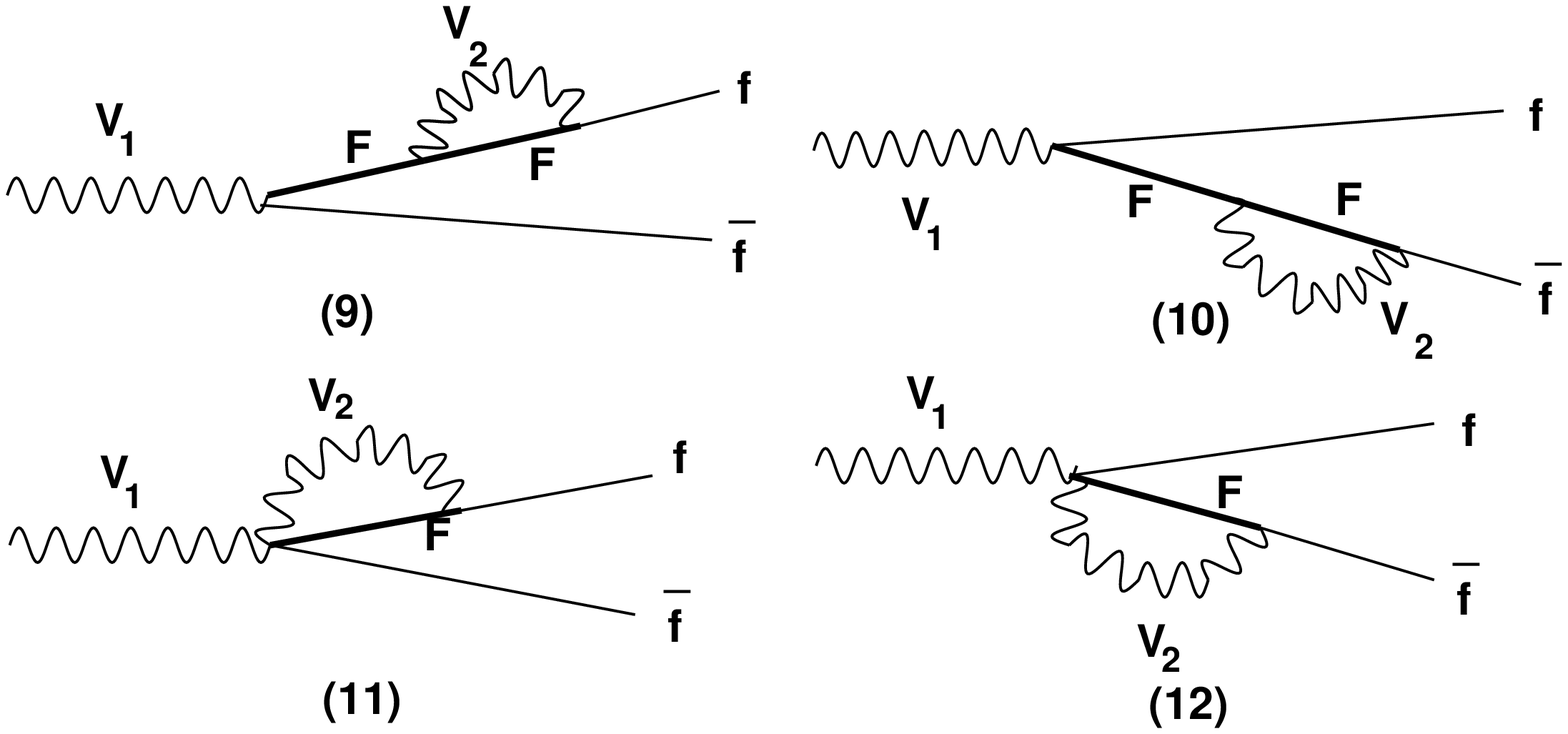}
\end{center}
\caption{The contribution of the excited leptons to the three-point
functions}
\label{fig:2}
\end{figure}

Corrections to the vertex $Z \bar{f} f$ give rise to
non-universal contributions to $\rho_f$ and $\kappa_f$.  Excited
leptons affect these couplings of the $Z$ through the diagrams
given in Fig.\  \ref{fig:2} whose results we parametrize as in
Ref.\ \cite{hollik},
\begin{equation}
-i \frac{e}{2 s_W c_W} \left [ \gamma_\mu F_{V{\text{ex}}}^{Zf} - 
\gamma_\mu \gamma_5
F_{A{\text{ex}}}^{Zf} -
 I_3^f \gamma_\mu (1 - \gamma_5) \frac{c_W}{s_W} ~
\frac{\Sigma^{\gamma Z}_{\text{ex}}(0)}{M_Z^2} \right ] \; ,
\end{equation}
where the singular part proportional to $\Sigma^{\gamma Z}
(0)$ has been split off, and
\begin{eqnarray}
\Delta \rho^{\text{ex}}_{\text{non}}  & = &  
\frac{2\, F_{A{\text{ex}}}^{Zf}(M_Z^2)}{I^f_3}
\; , \\
\Delta \kappa^{\text{ex}}_{\text{non}}  & = &  - \frac{1}{2 s_W^2  Q^f} 
\left [ F_{V{\text{ex}}}^{Zf} - 
\frac{I_3^f-2 s_W^2 Q^f}{I_3^f}~ F_{A{\text{ex}}}^{Zf}(M_Z^2)
\right ]
\; .
\end{eqnarray}
There are twelve one-loop Feynman diagrams that involve the
contribution of excited fermions to the three-point functions.
For each diagram we define $T^{V_2}_i (q^2, M^2, M_V^2)$, $i = 1,
\cdots , 12$, where $V_2$ is the virtual vector boson, with mass
$M_V$, running in the loop. In our calculations, we have
assumed that the ordinary fermions are massless ({\it i.e.\/}
$m^2 \ll M^2, M_V^2$), and in this limit, $T^{V_2}_{7, 8, 9, 10}
(q^2,M^2, M_V^2) = 0$.  Notice that the external fermion loops
(diagrams 5--10 of Fig.\ \ref{fig:2} ) only contribute as half,
due to the addition of the fermion wave function renormalization
counterterms. We also found the relations,
\begin{displaymath}
\begin{array}{l}
T^{V_2}_2(q^2, M^2, M_V^2) = T^{V_2}_3(q^2, M^2, M_V^2) \; ,
\\
T^{V_2}_5(q^2, M^2, M_V^2) = T^{V_2}_6(q^2, M^2, M_V^2) \; , 
\\
T^{V_2}_{11}(q^2, M^2, M_V^2) = T^{V_2}_{12}(q^2, M^2, M_V^2)
\end{array}
\end{displaymath}

Therefore, we can write the excited lepton contribution to the
form factors $F_{V(A){\text{ex}}}^{V_1f}$, for an external vector
boson $V_1$ as, 
\begin{equation}
F_{V{\text{ex}}}^{V_1f}(q^2)=F_{A{\text{ex}}}^{V_1f}(q^2)=
\frac{i s_W c_W}{e}
T_{V_1\rightarrow \bar{f} f} (q^2)\; ,
\label{mat}
\end{equation}
with
\begin{equation}
\begin{array}{ll}
T_{V_1\to f^+ f^-}(q^2) = & 
T_1^\gamma (q^2, M^2, 0) + T_1^Z (q^2, M^2, M_Z^2) + T_1^W (q^2, M^2, M_W^2) 
\\
&+
2\;\left[T_2^\gamma (q^2, M^2, 0) + T_2^Z (q^2, M^2, M_Z^2) + 
T_2^W (q^2, M^2, M_W^2) \right]\\
&+ T_4^W (q^2, M^2, M_W^2) 
\\
& +
\left[ T_5^\gamma (q^2, M^2, 0) + T_5^Z (q^2, M^2, M_Z^2) + 
T_5^W (q^2, M^2, M_W^2)
\right]
\\
& + 2\; T_{11}^W (q^2, M^2, M_W^2)\; .
\end{array}
\label{g:ee}
\end{equation}

Our results for $T^{V_2}_{1, 2, 4, 5, 11} (q^2, M^2, M_V^2)$,
in terms of the Passarino--Veltman scalar one-loop functions,
are
\begin{equation}
\begin{array}{ll}
T^{V_2}_1 = &
 \frac{\displaystyle i}{\displaystyle 4 \pi^2 \, q^2 }\,{{C^2_{V_2Ff}}}\,
\Biggl\{ \biggl[ A_{V_1FF}\, \Bigl( 2\,{M^6} - 3\,{M^4}\,{{M_V}^2} + 
{{M_V}^6} 
+ {M^2}\,{{M_V}^2}\,{q^2} + 2\,{{M_V}^4}\,{q^2}
\Bigr)
\\
& + K_{V_1FF}\, \left( 2\,M\,{{M_V}^4}\,{q^2} - 
2\,{M^3}\,{{M_V}^2}\,{q^2}\right) \biggr]
 \times \,  C_0 (0, 0, q^2, M^2, M_V^2, M^2)
\\
& + A_{V_1FF}\, 
\Bigl( -2\,{M^4} +
\,{M^2}\,{{M_V}^2} +  \,{{M_V}^4} 
+ \frac{\displaystyle 1}{\displaystyle 3}\,{M^2}\,{q^2}   
+ \frac{\displaystyle 2}{\displaystyle 9}\, q^4  
\Bigr)  +  K_{V_1FF}\,   \Bigl(\,{M}\,{q^4} +\, 2 M M_V^2 q^2 \Bigr) 
\\
& 
- \frac{\displaystyle (4\, M^2 - q^2)^{1/2}}{\displaystyle q}
\biggl[ A_{V_1FF}\, \Bigl(\, -12 M^4+\,6 M^2\, M_V^2 +\, 6 \, M_V^4 +\, 
10 \, M^2 \, Q^2 +\, 9 \, M_V^2 \, Q^2 -4 \, Q^4\Bigr) 
\\
& + K_{V_1FF}\,   \Bigl(\, 12 \, M \, M_V^2 \, Q^2 -6 \, M \,Q^4 \Bigr)\biggr] 
\, \times 
\arctan\left[\frac{\displaystyle q}
{\displaystyle (4\, M^2-q^2)^{1/2}}\right]
\\
& + \frac{\displaystyle q^2}{\displaystyle 6}\,
\left[ A_{V_1FF}\, \left( 18\,{M^2} + 9\,{{M_V}^2} - 
4\,{q^2}\right)  - 6\,K_{V_1FF}\,M\,{q^2} \right] \,
\log \frac{\displaystyle \Lambda^2}{\displaystyle M^2}\\
& 
- \frac{\displaystyle M_V^2}{\displaystyle (M^2-M_V^2)}
\,
\Bigl[
A_{V_1FF}\, \left( 2\,{M^4} - {M^2}\,{{M_V}^2} - 
{{M_V}^4} \right)
- 2\,K_{V_1FF}\,M\,{{M_V}^2}\,{q^2}
\Bigr] \,\log \frac{\displaystyle M^2}{\displaystyle M_V^2} \Biggr\}
\\
\;\;\;\; \simeq & 
 \frac{\displaystyle - i M^2}{\displaystyle 144 \pi^2 }\, C^2_{V_2Ff} \,
\Biggl\{  A_{V_1FF}(126+117 R_V) - R_Q \Bigl[ A_{V_1FF}(64 +9 R_V)
\\
& +K_{V_1FF}\,M (108+18 R_V) \Bigr] + \Bigl[ -A_{V_1FF}\, 
(108+54R_V) 
\\
&
+ R_Q \left( 24 \,A_{V_1FF}
+36 \,K_{V_1FF}\,M\right)\Bigr] \,
\log \frac{\displaystyle \Lambda^2}{\displaystyle M^2} \Biggr\}
\; , 
\end{array}
\label{t1:final}
\end{equation}
where the coupling constants $C_{VFf}$, $A_{VFF}$, and $K_{VFF}$
are given by (\ref{CV}), (\ref{AV}), and (\ref{KV}),
respectively, and the Passarino--Veltman function $C_0 (0, 0,
q^2, M^2, M_V^2, M^2)$ is given in  Appendix \ref{pave}. The
approximate expression was obtained for the large-$M$ limit, {\it
i.e.\/} $R_Q = q^2/M^2 \ll 1$ and  $R_V=M_V^2/M^2\ll 1$.
\begin{equation}
\begin{array}{ll}
T^{V_2}_2  = & 
 \frac{\displaystyle -i}{\displaystyle 4\pi^2}\,C_{V_1Ff}\,C_{V_2Ff}\,
        \left( g^a_{V_2} + g^v_{V_2} \right) \,\Biggl\{
        {M^2} - 2\,{{M_V}^2} - 2\,{q^2}  -\,{q^2}\,
        \log \frac{\displaystyle \Lambda^2}{\displaystyle M^2}  
        +2 {{M_V}^2}  \,
        \log \frac{\displaystyle M^2}{\displaystyle M_V^2}
\\
&+2 {{M_V}^2}\,
        \left( {M^2} - {{M_V}^2} - {q^2} \right)\,
C_0 (0, 0, q^2, M^2, M_V^2, 0) \\
& +  \frac{\displaystyle M^2 - q^2}{\displaystyle q^2}\,
        \left( {M^2} - 2\,{{M_V}^2} - {q^2} \right) \,
\log \left(1 - \frac{\displaystyle q^2}{\displaystyle M^2} \right )\Biggr\} 
\\
\;\;\;\; \simeq & 
\frac{\displaystyle i M^2 }{\displaystyle 8\pi^2}\,C_{V_1Ff}\,C_{V_2Ff}\,
        \left( g^a_{V_2} + g^v_{V_2} \right) \,
R_Q \left( 1+2 R_V \log R_V +
2  \log \frac{\displaystyle \Lambda^2}{\displaystyle M^2} 
\right)
\; , 
\end{array}
\label{t2:final}
\end{equation}
where $g^v_{V}$ and $g^a_{V}$ are the vector and axial coupling of the
vector bosons to the usual fermions: 
for $V = \gamma$, $g^v_{\gamma} = - e$  and $g^a_{\gamma} = 0$;  
for $V = W$,      $g^v_{W} =  g^a_{W} =  g/(2 \sqrt{2})$; 
for $V = Z$  and  $f = \nu$, $g^v_{Z} =  g^a_{Z} = g/(4 c_W)$;
for $V = Z$  and  $f = e$, $g^v_{Z} = g (4 s_W^2 - 1)/( 4
c_W)$  and  $g^a_{Z} = - g/(4 c_W)$. 

\begin{equation}
\begin{array}{ll}
T^{V_2}_4 = &  
\frac{\displaystyle i}{\displaystyle 144\pi^2\,q^2} 
\, C^2_{V_2Ff}\, g_{V_1WW}\,
        \Biggl\{ - 36\, \Lambda^2\, q^2 + 72\, M^4  - 36\, M^2 \,M_V^2
- 36\, M_V^4 - 45\, M^2 \, q^2  
\\
& 
+ 15\, M_V^2\, q^2 + 46\, q^4 
+ 18 \Bigl( 4\, M^6 - 6\, M^4\,M_V^2 
+ 2\, M_V^6 - M^4\, q^2 + 4\, M^2\,M_V^2\, q^2 
\\
& + 3\,M_V^4\, q^2  - M^2\,q^4 \Bigr) \times \, 
C_0 (0, 0, q^2, M_V^2, M^2, M_V^2) 
\\
&- 6 \frac{\displaystyle (4\, M_V^2 - q^2)^{1/2}}{\displaystyle q}
\,
        \Bigl( 24\, M^4 - 12\,M^2\, M_V^2 - 12\, M_V^4 
- 18\, M^2 \, q^2  + 4\, M_V^2 \, q^2 
\\
&  + 5\, q^4 \Bigr) \, \times 
\arctan\left[\frac{\displaystyle q}
{\displaystyle (4\, M_V^2-q^2)^{1/2}}\right]
+ 3\,q^2\,\left( 18\, M^2  + 36\, M_V^2 + 5\, q^2 \right)
\,\log \frac{\displaystyle \Lambda^2}{\displaystyle M^2}
\\
&+\frac{\displaystyle 3}{\displaystyle(M^2 - M_V^2)} 
\Bigl( 24\, M^6 - 12\, M^4 \, M_V ^2  - 12\, M^2 \, M_V^4  
- 18\, M^4 \, q^2  -  36\, M_V^4 \, q^2
\\
&  + 5\, M^2 \, q^4  - 5\, M_V^2\, q^4 \Bigr) \,
\log \frac{\displaystyle M^2}{\displaystyle M_V^2} \Biggr\} 
\\
\;\;\;\; \simeq & 
\frac{\displaystyle i M^2}{\displaystyle 288\pi^2} 
\, C^2_{V_2Ff}\, g_{V_1WW}\,
 \Biggl[ -72 \frac{\displaystyle\Lambda^2}{\displaystyle M^2} -
 18-36 R_V +R_Q 
\left( 103 +144 R_V+144 R_V \log R_V \right) \\
& +\left( 108+216 R_V +30 R_Q\right)  
\log \frac{\displaystyle \Lambda^2}{\displaystyle M^2}\Biggr]
\; , 
\end{array}
\label{t4:final}
\end{equation}
where $g_{V_1WW}$  is the coupling constant of the triple vector
boson vertex. For $V_1 = \gamma, Z$ is given by  $g_{\gamma WW} =
g s_W$  and $g_{ZWW} = g c_W$.

\begin{equation}
\begin{array}{ll}
T^{V_2}_5 = & 
\frac{\displaystyle i}{\displaystyle 16\left(M^2 - M_V^2 \right) \,\pi^2}
\,{{C^2_{V_2Ff}}}\,\left( g^a_{V_1} + g^v_{V_1} \right) \,
   \Biggl[ 14\,{M^4} - {M^2}\,{{M_V}^2} - 7\,{{M_V}^4} 
\\
&- 6(M^2-M_V^2)\,
      \left( 2\,{M^2} + {{M_V}^2} \right) \,
      \log \frac{\displaystyle \Lambda^2}{\displaystyle M^2} 
- 6 \frac{\displaystyle M_V^6}{\displaystyle  M^2-M_V^2}\,
  \log \frac{\displaystyle M^2}{\displaystyle M_V^2}
\Biggr]
\\
\;\;\;\; \simeq &
\frac{\displaystyle i M^2}{\displaystyle 16\,\pi^2}
\,{{C^2_{V_2Ff}}}\,\left( g^a_{V_1} + g^v_{V_1} \right) \,
   \Biggl[ 14+13 R_V-6 (2+ R_V) 
\log \frac{\displaystyle \Lambda^2}{\displaystyle M^2} \Biggr]\; , 
\end{array}
\label{t5:final}
\end{equation}
and,
\begin{equation}
\begin{array}{ll}
T^{V_2}_{11} =&  
\frac{\displaystyle - i}{\displaystyle 32\pi^2}
\,C_{V_2Ff}\,D_{V_1Ff}\,\Biggl[
      4\,{{\Lambda}^2} + 15\,{M^2} + 15\,{{M_V}^2} 
- 18\left( M^2 + M_V^2 \right) \,
      \log \frac{\displaystyle \Lambda^2}{\displaystyle M^2}
\\
&+ \frac{\displaystyle 18 M_V^4}{\displaystyle M^2-M_V^2}\,
      \log \frac{\displaystyle M^2}{\displaystyle M_V^2} \Biggr] \\
\;\;\;\; \simeq & 
\frac{\displaystyle -i M^2}{\displaystyle 32\pi^2}
\,C_{V_2Ff}\,D_{V_1Ff}\,
\Biggl[ 4 \frac{\displaystyle \Lambda^2}{\displaystyle M^2} + 15\,
+ 15 R_V - 18(1+R_V)  
\log \frac{\displaystyle \Lambda^2}{\displaystyle M^2} \Biggr]
\; , 
\end{array}
\label{t11:final}
\end{equation}
where $D_{VFf}$ is given in (\ref{DV}).

In order to make a consistency check of the whole calculation, we
have  analyzed the effect of the excited leptons to the $\gamma
\bar{f} f$ vertex at zero momentum, which is used as one of the
renormalization conditions in the on-shell renormalization
scheme. Taking into account the appropriate values for the
constants $C_{VFf}$ (\ref{CV}),  $A_{VFF}$ (\ref{AV}), and
$K_{VFF}$ (\ref{KV}), we verified that our exact result, {\it
c.f.\/} Eq.\ \ref{g:ee}, for the vertex $\gamma \bar{e} e$
cancels at $q^2=0$. This result should be expected since we are
using a gauge invariant effective Lagrangian, and the QED Ward
identities \cite{ward} require that the excited fermion
contribution to this vertex, at zero momentum, vanishes. In the
same way, we have also checked that $T_{\gamma \to \bar{\nu}
\nu}(q^2 = 0) = 0$ (note that $T_4$ and $T_{11}$ must change sign
for external neutrinos). Moreover, we also verified that the
excited fermions decouple from the vertex correction in the limit
of large $M$.

It should be pointed out the the high energy cutoff of the loop
integrals represents the maximum energy to which the effective
Lagrangian is expected to apply and  we have  assumed that the
effective operators are valid just up to the compositeness scale,
$\Lambda$.  Therefore, decoupling of heavy excited states occurs
only when $M \to \infty$ while keeping the ratio $M/\Lambda$
finite. 

Finally, we present an approximated expression for the form factor
$F_{V{\text{ex}}}^{V_1f}$, at first order in $R_Q$, $R_Z$ and
$R_W$, which is valid for $V_1=\gamma, Z$:
\begin{equation}
\begin{array}{ll}
F_{V{\text{ex}}}^{V_1f}(q^2) =& 
-\frac{\displaystyle  M^2 s_W c_W}{ \displaystyle 288 \pi^2  e}\, R_Q \,
      \Biggl\{ 128\,A_{V_1 F'F'}\, C^2_{WF'f} + 
        128\,A_{V_1 FF}\,\left( C^2_{\gamma Ff} + C^2_{ZFf}\right) 
\\
&     + 72\, C_{V_1F'f'}\, C_{WF'f}\, \left( g^a_W + g^v_W \right)  + 
        72\,C_{V_1Ff}\,\left[ C_{\gamma Ff}\,(g^a_{\gamma f} +g^v_{\gamma f}) 
        + C_{ZFf}\,  (g^a_{Z f} +g^v_{Z f}) \right] 
\\
&       +103\, C^2_{WF'f}\, g_{V_1WW} + 
        216\,\left( C^2_{\gamma Ff} \, K_{V_1FF} + 
           C^2_{ZFf} \, K_{V_1FF} +  C^2_{WF'f}\, K_{V_1F'F'} \right)
          \,M  
\\
&      +  \Biggl[ 48\,A_{V_1 F'F'}\, C^2_{WF'f} + 
           48\,A_{V_1 FF}\,\left( C^2_{\gamma Ff} + C^2_{ZFf} \right) 
            - 144\, C_{V_1F'f'}\, C_{WF'f}\,
            \left( g^a_W + g^v_W \right) 
\\
&      -  144\,C_{V_1Ff}\,\left[ C_{\gamma Ff}\, 
            (g^a_{\gamma f} +g^v_{\gamma f}) + 
              C_{ZFf}(g^a_{Z f}+ g^v_{Z f})\right]  - 
           30\, C^2_{WF'f}\, g_{V_1WW} 
\\
&         +72\,\left( C^2_{\gamma Ff}\, K_{V_1FF} + 
               C^2_{ZFf} \,K_{V_1FF} +  C^2_{WF'f}\, K_{V_1F'F'} 
                   \right) \,M  \Biggr] \,\log R_L 
\\
&        + R_W\,\Bigl[ 18\,A_{V_1 F'F'}\, C^2_{WF'f} + 
           144\, C^2_{WF'f}\, g_{V_1WW} + 
           36\, C^2_{WF'f}\,K_{V_1F'F'}\,M 
\\
&         +144\, C_{WF'f}\,\left[ C_{V_1F'f'}\, ( g^a_W + g^v_W)
           +C_{WF'f}\, g_{V_1WW} \right]\log R_W \Bigr]
\\
 &       + R_Z \,\Bigl[ 18\,A_{V_1 FF}\, C^2_{ZFf} + 
           36\, C^2_{ZFf}\,K_{V_1FF}\,M 
\\
&
+ 144\,C_{V_1Ff}\, C_{ZFf}\,
            \left( g^a_{Zf} + g^v_{Zf} \right) \,\log R_Z \Bigr]  
          \Biggr\}\; .
\end{array}
\label{vertex}
\end{equation}

From the above equation it is evident that the vertex corrections
are proportional to $R_Q$, and therefore vanishes at $q^2 = 0$.

\section{Numerical Results and Discussion}
\label{sec:disc}

\begin{figure}
\protect
\epsfxsize=14cm
\begin{center}
\leavevmode \epsfbox{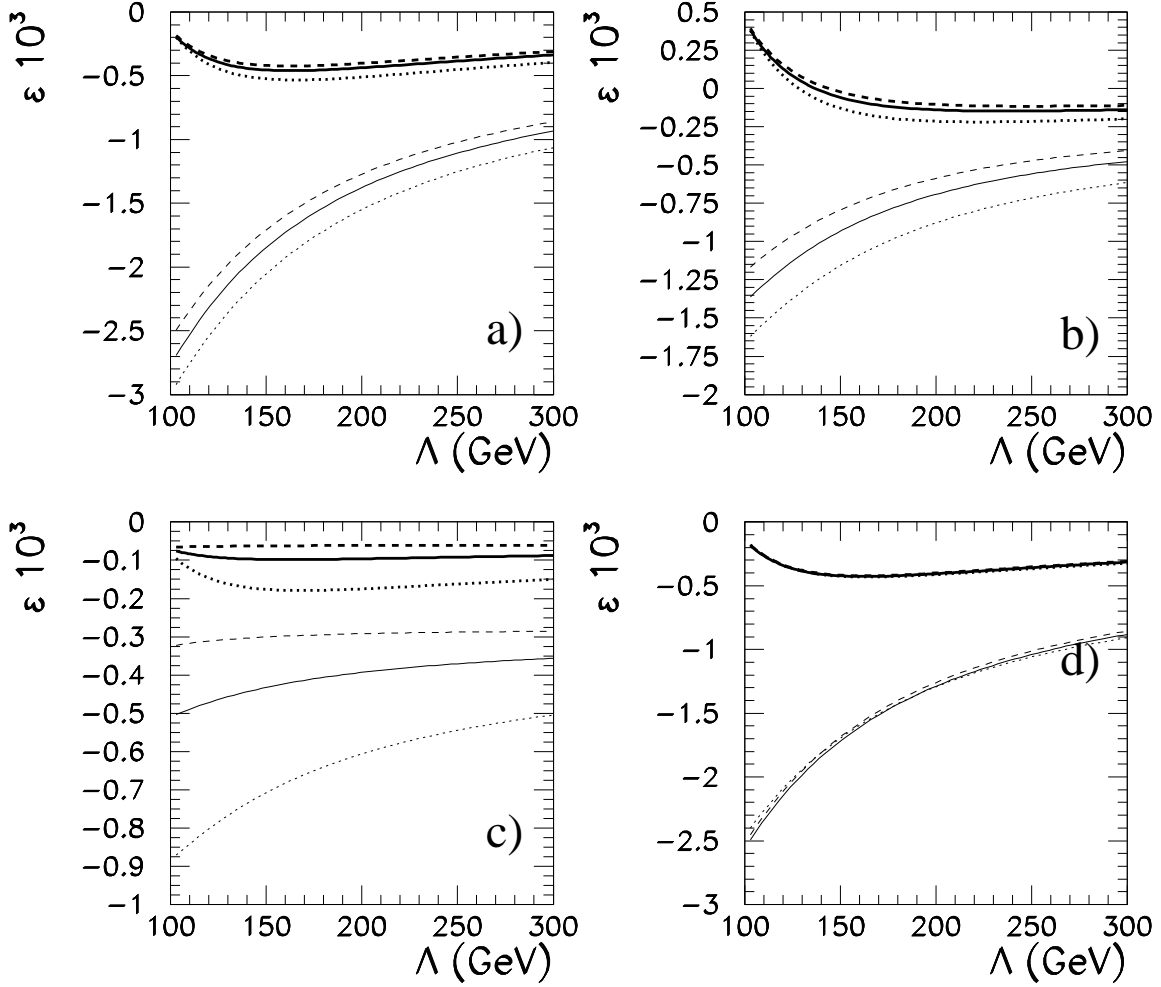}
\end{center}
\caption{Attainable values for the new contributions to the
$\epsilon$'s  parameters in the model as a function of the scale
$\Lambda$. The solid lines correspond to $\epsilon_1$, the dashed
ones to $\epsilon_2$ and the dotted ones to  $\epsilon_3$. The
thin (thick) lines correspond to excited lepton mass  value of
$M=100$ (200) GeV. We have assumed different configurations of
the weight factors \protect{$(f_1,f_2,\kappa_1,\kappa_2)$}: 
(a) = $(1,1,1,1)$; (b) = $(1,-1,1,-1)$; 
(c) = $(1,0,1,0)$; (d) = $(0,1,0,1)$ }
\label{fig:3}
\end{figure}

The above expressions for the radiative corrections to $Z$
physics due to excited leptons are valid for arbitrary couplings
and masses. In order to gain some insight as to  which
corrections are the most relevant, let us begin our analyses by
studying just the oblique corrections, which can also  be
parametrized in terms of the variables $\epsilon_1$,
$\epsilon_2$, and $\epsilon_3$ of Ref.\ \cite{eps}
\begin{equation} 
\begin{array}{l}
\epsilon^{\text{ex}}_1 = \Delta \rho^{\text{ex}}_{\text{univ}}(z)\\   
\epsilon^{\text{ex}}_2 =
c_W^2\Delta \rho^{\text{ex}}_{\text{univ}}(z) -2 s_W^2 \Delta
\kappa^{\text{ex}}_{\text{univ}}-s_W^2 \Delta r^{\text{ex}}_{\text{univ}}\\ 
\epsilon^{\text{ex}}_3
=c_W^2\Delta \rho^{\text{ex}}_{\text{univ}}(z) +c_W^2\Delta
r^{\text{ex}}_{\text{univ}}+(c_W^2-s_W^2)
\Delta \kappa^{\text{ex}}_{\text{univ}}(z)\; .
\end{array} 
\label{eps} 
\end{equation} 
Recent global analyses of the LEP, SLD, and low-energy data yield
the following values for the oblique parameters
\cite{eps} 
\begin{equation} 
\begin{array}{l} 
\epsilon_1 =\epsilon_1^{\text{SM}} + \epsilon_1^{\text{new}} 
= (5.1\pm 2.2)\times 10^{-3} 
\; ,
\\ \epsilon_2 = \epsilon_2^{\text{SM}} + \epsilon_2^{\text{new}} 
= (-4.1\pm
4.8)\times 10^{-3} \; , \\ 
\epsilon_3= \epsilon_3^{\text{SM}} + \epsilon_3^{\text{new}}= 
(5.1\pm 2.0)\times 10^{-3} \; ,
\end{array} 
\label{epsi}  
\end{equation} 
In Fig \ref{fig:3}, we give the attainable values for the
new contributions to the $\epsilon$  parameters  for different
values of the excited lepton mass and couplings.  As seen from
this figure, requiring that the new contribution is within the
limits allowed by the experimental data (\ref{epsi}), we find
that the constraints coming from oblique corrections are less
restrictive than the available experimental limits. Notice that
$\Lambda$ being the scale of new physics, $M$ must satisfy $M\leq
\Lambda$.

As for the vertex corrections, we see in Eq. (\ref{mat}) that the
excited leptons alter just the left-handed-lepton coupling of the
$Z$.  The new  contributions to the $Z$ widths, $\Gamma_{ee}
\equiv \Gamma (Z \to e^+ e^-)$ and $\Gamma_{\text{inv}} \equiv 3
\; \Gamma(Z \to \bar{\nu} \nu)$, are given by
\begin{equation}
\begin{array}{l}
\Delta \Gamma_{ee} =  \alpha M_Z \frac{\displaystyle (s_W^2 - 1/2)}
{\displaystyle 3 s_W^2 c_W^2 } \times  F^{Ze}_{V}(z) \; ,
\nonumber \\
\Delta \Gamma_{\text{inv}}  =  \alpha  M_Z \frac{\displaystyle 1}
{\displaystyle 2 s_W^2 c_W^2 }\times   F^{Z\nu}_{V}(z)  \; .
\end{array}
\label{z:wid}
\end{equation}

The theoretical values for the $Z$ partial width generated by
ZFITTER \cite{zfit}, for $m_{\text{top}} = 175$ GeV and $M_H =
300$ GeV, are $\Gamma_{ee} = 83.9412$ MeV and
$\Gamma_{\text{inv}} = 501.482$ MeV.  The most recent LEP results
\cite{hep:conf}, assuming lepton universality, are
$\Gamma_{\ell\ell}^{\text{LEP}} (Z \to \ell^+ \ell^-) = 83.93 \pm
0.14$ MeV and for the  invisible width
$\Gamma_{\text{inv}}^{\text{LEP}} = 499.9 \pm 2.5$ MeV.
Therefore, at $95\%$ C.L., we should have $-0.28 < \Delta
\Gamma_{ee} < 0.26$ MeV, and $- 6.48 < \Delta \Gamma_{\text{inv}}
< 3.32$ MeV.

\begin{figure}
\protect
\epsfxsize=15cm
\begin{center}
\leavevmode \epsfbox{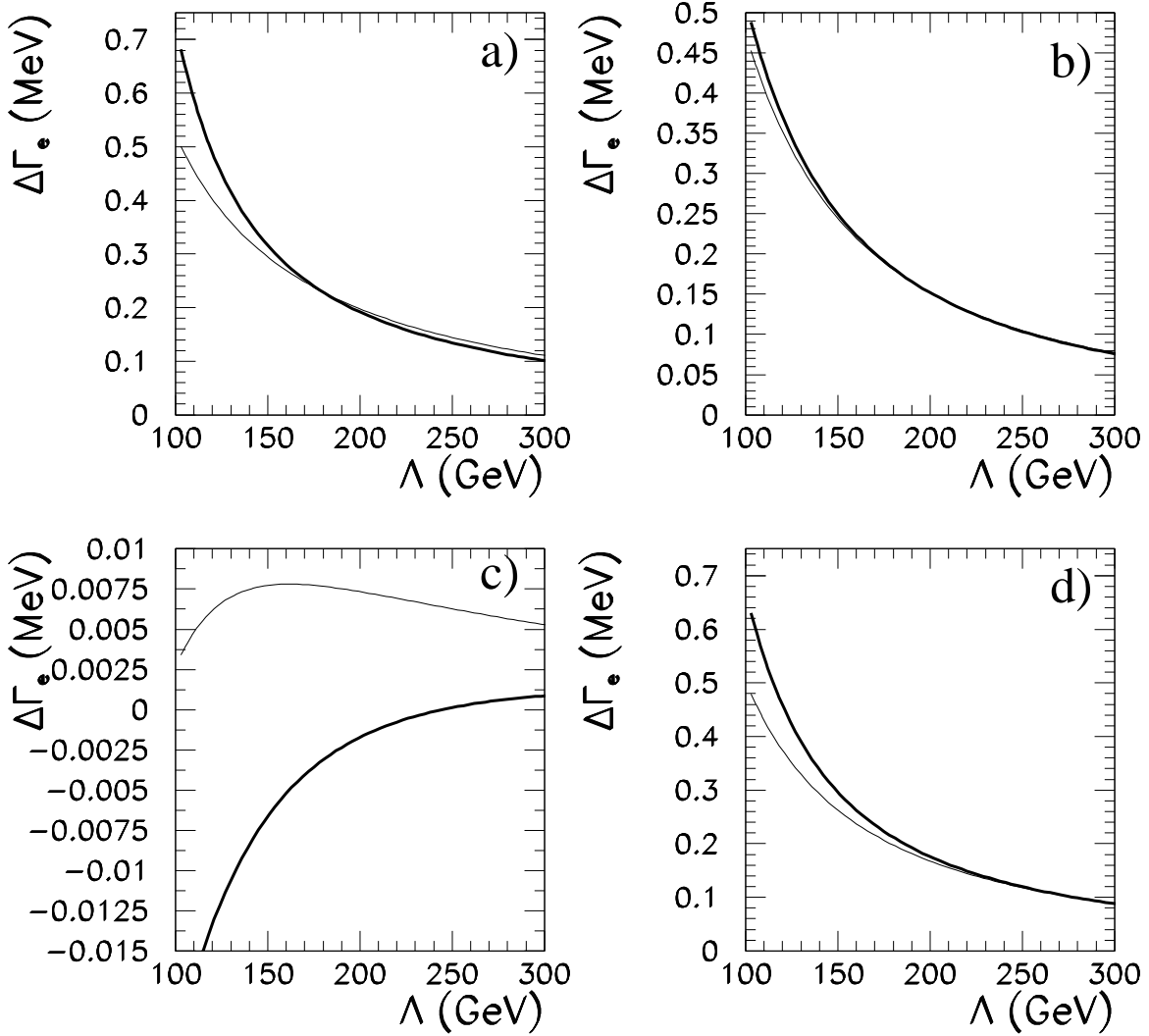}
\end{center}
\caption{Attainable values for the new contributions to the width 
$\Gamma(Z\rightarrow \ell^+ \ell^-)$ in the model as a function of the scale
$\Lambda$. The thin (thick) line correspond to excited lepton mass value
of $M=100$ (200) GeV, for configurations of the weight factors as
in Fig.\ \protect{\ref{fig:3}}.}
\label{fig:4}
\end{figure}

\begin{figure}
\protect
\epsfxsize=15cm
\begin{center}
\leavevmode \epsfbox{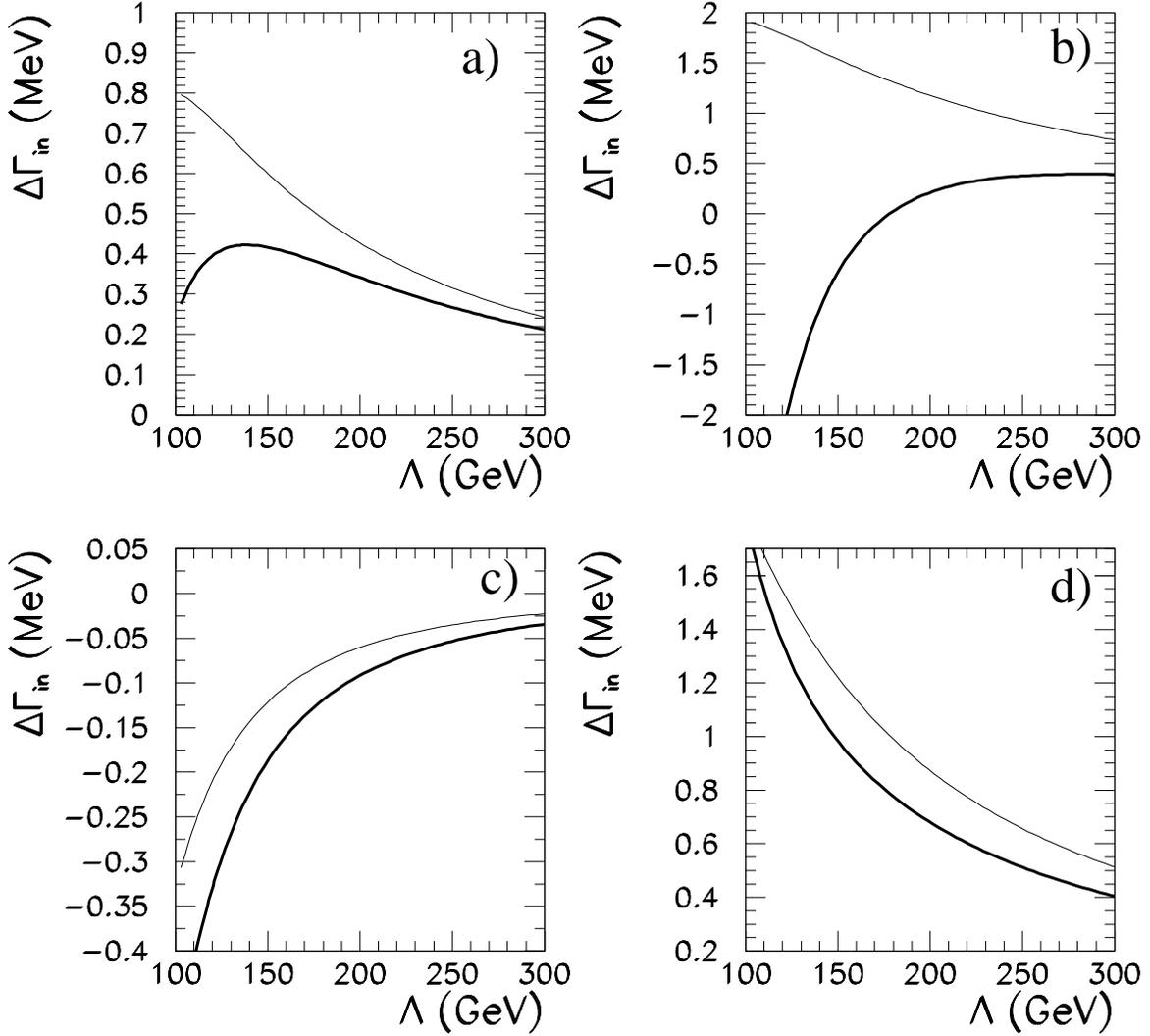}
\end{center}
\caption{Attainable values for the new contributions to the
invisible $Z$ width in the model as a function of the scale
$\Lambda$. The thin (thick) line correspond to excited lepton
mass value of $M=100$ (200) GeV, for configurations of the weight
factors as in Fig.\ \protect{\ref{fig:3}}.}
\label{fig:5}
\end{figure}

We plot in Figures \ref{fig:4} and \ref{fig:5}  the
values of  $\Delta\Gamma_{ee}$ and $\Delta\Gamma_{\text{inv}}$
attainable in this phenomenological model for some values of the
compositeness scale and excited lepton masses, assuming different
configurations of the weight factors $f_{1,2}$, and
$\kappa_{1,2}$. Our numerical results show that the most
restrictive bound on the excited fermion mass and compositeness
scale comes from the comparison of $\Delta \Gamma_{ee}$ with the
LEP data for this observable. 

Let us compare our bounds coming from $\Delta \Gamma_{ee}$ with
the ones emerging from the direct search for the excited leptons.
First of all, we should point out that the direct search at LEP
was  just able to reach excited fermion masses up to 130 GeV
\cite{lep:130}. On the other hand, the HERA Collaborations
\cite{hera,zeus}, looking for states that decay into a gauge
boson and a usual fermion in the reaction $e p \to f^\ast X$, can
access masses up to 250 GeV. 

\begin{figure}
\protect
\epsfxsize=15cm
\begin{center}
\leavevmode \epsfbox{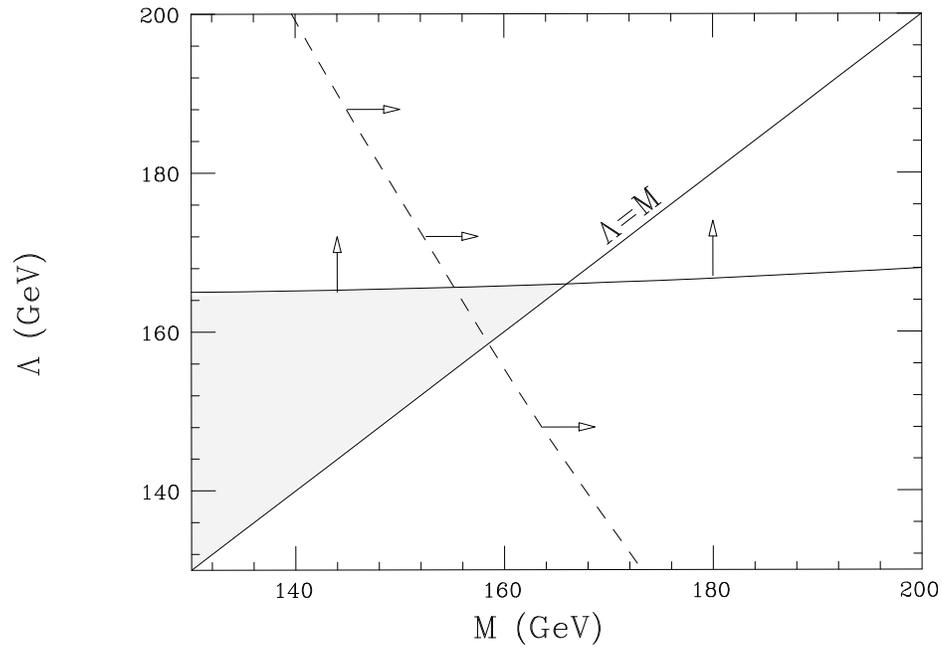}
\end{center}
\caption{Excluded regions in the $\Lambda$ versus  $M$ plane from the
bounds on \protect{$\Delta \Gamma_{ee}$} (shadowed area),  
and from ZEUS data \protect{\cite{zeus}} (below and
left of the dashed curve), at \protect{$95\%$}  C.L.}
\label{fig:6}
\end{figure}

In Fig.\ \ref{fig:6}, we present the excluded region, at $95\%$
C.L., in the $\Lambda$ versus  $M$ plane imposed by $\Delta
\Gamma_{ee}$, for $f_1 = f_2 = \kappa_1 = \kappa_2 = 1$. We have
further assumed that $M \leq \Lambda$, leading to the excluded
region represented by the shadowed triangle. For comparison, we
also present the region excluded by the ZEUS data \cite{zeus}
(below and left of the dashed curve), for  $f_1 = f_2 = 1$. Since
we have assumed that $BR(e^\ast \to e \gamma) = 1$, this curve
represents an upper limit for the ZEUS bound. As we can see, we
were able to exclude just a small region beyond the available
limit.  We also show our results when  we relax the condition of
$M \leq \Lambda$. In the latter case, our analysis  excludes all
excited lepton masses with scales $\Lambda \leq 165$ GeV.

In principle, compositeness may not only generate these operators
involving excited leptons which contribute to vector boson self
energies and vertices at one-loop level, but it may also generate
effective operators which could give tree-level contributions
which we did not consider. Cancellations could then be possible
between tree-level and one-loop contributions and the bounds
derived here would not be applied. We assumed that it is
unnatural that large cancellations occur between the tree-level
and the one-loop contributions in the observables measured at
$M_Z$ scale \cite{zep}.

In conclusion, we have evaluated the contribution of excited
lepton states, up to the one-loop level, to the oblique variables
and also to the $Z$ width to leptons. We have compared our
results with the precise data on the electroweak observables
obtained by the LEP Collaborations in order to extract bounds on
some of the free parameters (compositeness scale and excited
lepton mass) of the phenomenological model under consideration.
We also compared our results with the recent bounds obtained
through the direct search for these particles. Our results show
that the present precision in the electroweak parameters attained
by LEP is very marginally able to  constrain the parameters
$\Lambda$ and $M$ beyond the present limits from direct searches. 

\acknowledgments

One of us (S.F.N.) would like to thank the Instituto de
F\'{\i}sica  Corpuscular -- IFIC/C.S.I.C.\ of Universidad de
Val\`encia for its kind  hospitality. We would like to thank Dr.\
R.\ Mertig for providing us with the last version of the FeynCalc
package, and Dr.\ O. J.\ P.\ \'Eboli for useful discussions. This
work was partially supported by Generalitat Valenciana  (Spain),
and Conselho Nacional de Desenvolvimento Cient\'{\i}fico e
Tecnol\'ogico (Brazil).

\newpage
\appendix
\section{Scalar One-Loop Integrals}
\label{pave}

The relevant Passarino--Veltman functions are \cite{passa}, 
\begin{eqnarray}
A_0(m_0^2) &=& -i (16 \pi^2) \mu^{4 - D} \int \frac{d^D k}{(2\pi)^D}
\frac{1}{k^2 - m_0^2} \; ,
\nonumber \\
B_0(p_1^2, m_0^2, m_1^2) &=& -i (16 \pi^2) \mu^{4 - D} 
\int \frac{d^D k}{(2\pi)^D}
\frac{1}{(k^2 - m_0^2) [(k + p_1)^2 - m_1^2]} \; ,
\\
\label{ABC}
C_0 (p_{1}^2, p_{21}^2, p_{2}^2, m_0^2, m_1^2, m_2^2) &=& 
-i (16 \pi^2) \mu^{4 - D} \int \frac{d^D k}{(2\pi)^D}
\frac{1}{(k^2 - m_0^2) [(k + p_1)^2 - m_1^2] [(k + p_2)^2 - m_2^2]} 
\nonumber
\end{eqnarray}
where $p_{21} = p_2 - p_1$

The scalar function $A_0$ can be written as\cite{den},
\begin{equation}
A_0(m_0^2) = m_0^2 \left( \Delta - \log\frac{m_0^2}{\mu^2} + 1 \right) +
2\frac{\mu^2}{\pi (D-2)}
\label{A0}
\end{equation}
where we have kept the pole at $D=2$, and 
\begin{equation}
\Delta = \frac{2}{4-D} - \gamma_E + \log 4\pi
\label{delta}
\end{equation}
where $\gamma_E$ is Euler's constant. 

The $B_0$ and $C_0$ functions can be written in terms of
integrals over Feynman parameters as
\begin{equation}
B_0(p_1^2, m_0^2, m_1^2) = \Delta  - \int_0^1 dx 
\log\left[
\frac{x^2 p_1^2 - x (p_1^2 + m_0^2 - m_1^2) + m_0^2 - 
i \epsilon}{\mu^2}\right]
\label{B0}
\end{equation}
and
\begin{eqnarray}
C_0 (p_{1}^2, p_{21}^2, p_{2}^2, m_0^2, m_1^2, m_2^2) &=&
- \int_0^1 dx \int_0^x dy  
\Bigl[ p_{21}^2 x^2 + p_{1}^2 y^2 + (p_{2}^2 - p_{1}^2 -
p_{21}^2) x y 
\nonumber \\
&+& (m_1^2 -
m_2^2 - p_{21}^2) x + (m_0^2 - m_1^2 + p_{21}^2 - p_{2}^2) y +
m_2^2 - i \epsilon \Bigr]^{-1}
\label{C0}
\end{eqnarray}

The function $B_0$, for some cases of interest,  are 
\begin{equation}
\begin{array}{l}
B_0(0, 0, M^2) =  \Delta + 1 - 
\log\left(\frac{\displaystyle M^2}{\displaystyle \mu^2}\right)
\; ,
\\
B_0(0,M^2,M_V^2) = \Delta + 1 - 
\frac{\displaystyle (M^2 + M_V^2)}{\displaystyle 2 (M^2 -M_V^2)} 
\log\left(\frac{\displaystyle M^2}{\displaystyle M_V^2}\right) - 
\log\left( \frac{\displaystyle M M_V}{\displaystyle \mu^2}\right)\; ,
 \\
B_0(q^2,0,M^2) = \Delta + 2 - 
\left(1 - \frac{\displaystyle M^2}{\displaystyle q^2} \right)
 \log\left(1-\frac{\displaystyle q^2}{\displaystyle M^2}\right) 
- \log\left(\frac{\displaystyle M^2}{\displaystyle \mu^2}\right)
\; ,
\\
B_0(q^2, M^2, M^2) =  \Delta + 2 - 
2 \frac{(\displaystyle 4 M^2-q^2)^{1/2}}{\displaystyle q}
 \arctan \left[ \frac{\displaystyle q} {\displaystyle (4\, M^2 - q^2)^{1/2} }
\right]
-\log \frac{\displaystyle M^2}{\displaystyle \mu^2}
\end{array}
\label{B0:final}
\end{equation}

The functions $C_0$, for some cases of interest, are
\begin{equation}
\begin{array}{ll}
C_0 (0, 0, 0, M^2, 0, M^2) =& - \frac{\displaystyle 1}{\displaystyle M^2}
\; ,
\\
C_0 (0, 0, 0, M^2, M_V^2, 0) =& 
- \frac{\displaystyle 1}{\displaystyle( M^2 - M_V^2)}
\log\left(\frac{\displaystyle M^2}{\displaystyle M_V^2}\right)
\; ,
\\
C_0 (0, 0, 0, M^2, M_V^2, M^2) =& 
- \frac{\displaystyle 1}{\displaystyle (M^2 - M_V^2)^2}
 \left\{ M^2 - M_V^2 \left[ 1 + 
\log\left(\frac{\displaystyle M^2}{\displaystyle M_V^2}\right)
\right] \right\}
\; ,
\\
C_0 (0, 0, q^2, M^2, M_V^2, 0) =& \frac{\displaystyle 1}{\displaystyle q^2} 
\Biggl[
 \log\left(\frac{\displaystyle M^2 - q^2}{\displaystyle M_V^2}\right)
   \log\left( \frac{\displaystyle M^2 - q^2}{\displaystyle M_V^2} - 1 \right) 
\\
&- \log\left(\frac{\displaystyle M^2}{\displaystyle M_V^2} - 1\right)
     \log\left(\frac{\displaystyle M^2}{\displaystyle M_V^2}\right) 
\\
&+ i \pi \log\left(1 - \frac{\displaystyle q^2}{ \displaystyle M^2}\right) - 
    \text{Li}_2 \left(\frac{\displaystyle M^2}{\displaystyle M_V^2}\right)  + 
     \text{Li}_2 \left( \frac{\displaystyle M^2 - q^2}{\displaystyle M_V^2}
\right) \Biggr] \; ,
\\
C_0 (0, 0, q^2, M^2, M_V^2, M^2) =& \frac{\displaystyle 1}{\displaystyle q^2} 
\Biggl\{
-2 \pi 
\arctan\left[\frac{\displaystyle q (4 M^2 - q^2)^{1/2}}
{\displaystyle 2 (M^2 - M_V^2) -q^2}\right] 
\\
& + 4 
\arctan\left[\frac{\displaystyle (4 M^2 - q^2)^{1/2}}
{\displaystyle q} \right] 
\arctan\left[\frac{\displaystyle q (4 M^2 - q^2)^{1/2}}
{\displaystyle 2 (M^2 - M_V^2) - q^2}\right]  
\\
& -
\log\left(\frac{\displaystyle M^2}{\displaystyle M_V^2}\right) 
\log\left[\frac{\displaystyle (M^2 - M_V^2)^2}
{\displaystyle (M^2 - M_V^2)^2 + M_V^2 q^2}\right] 
\\
& - 
\text{Li}_2 \left(\frac{\displaystyle M^2 q^2}
{\displaystyle(M^2 - M_V^2)^2 + M_V^2 q^2}\right) + 
\text{Li}_2 \left(\frac{\displaystyle M_V^2 q^2}
{\displaystyle (M^2 - M_V^2)^2 + M_V^2 q^2}\right) 
\\
&+ \text{Li}_2 \left(\frac{ \displaystyle \xi^\ast}
{\displaystyle 2 (M^2 - M_V^2) - \xi}\right) -
\text{Li}_2 \left(\frac{\displaystyle - \xi^\ast}
{\displaystyle 2 (M^2 - M_V^2) - \xi^\ast}\right) 
\\
&+
\text{Li}_2 \left(\frac{ \displaystyle \xi}
{\displaystyle 2 (M^2 - M_V^2) - \xi^\ast}\right) - 
\text{Li}_2 \left( \frac{\displaystyle -\xi}
{\displaystyle 2 (M^2 - M_V^2) - \xi}\right)
\Biggr\} \; ,
\\
C_0 (0, 0, q^2, M^2, 0, M^2) =& \frac{\displaystyle 1}{\displaystyle q^2} 
\Biggl\{
-2 \pi 
\arctan\left[\frac{\displaystyle q (4 M^2 - q^2)^{1/2}}
{\displaystyle 2 M^2  -q^2}\right] -
\text{Li}_2 \left(\frac{\displaystyle M^2 q^2}
{\displaystyle M^4 }\right)  
 \\
& + 4 \arctan\left[\frac{\displaystyle (4 M^2 - q^2)^{1/2}}
{\displaystyle q} \right] 
\arctan\left[\frac{\displaystyle q (4 M^2 - q^2)^{1/2}}
{\displaystyle 2 M^2 - q^2}\right]  
\\
&+ \text{Li}_2 \left(\frac{ \displaystyle \xi^\ast}
{\displaystyle 2 M^2  - \xi}\right) -
\text{Li}_2 \left(\frac{\displaystyle - \xi^\ast}
{\displaystyle 2 M^2  - \xi^\ast}\right) 
+
\text{Li}_2 \left(\frac{ \displaystyle \xi}
{\displaystyle 2 M^2 - \xi^\ast}\right) \\
& -\text{Li}_2 \left( \frac{\displaystyle -\xi}
{\displaystyle 2 M^2 - \xi}\right)
\Biggr\} 
\; ,
\end{array}
\label{C0:final}
\end{equation}
where $\xi = q^2 + i q (4 M^2 - q^2)^{1/2}$, and $\text{Li}_2 (x)$
is the dilogarithm or Spence's function, defined as
\[
\text{Li}_2 (x) =  - \int_0^1 \frac{dt}{t}  \log(1 - xt) \; ,
\]



\end{document}